\documentclass[letterpaper,10pt,conference]{ieeeconf}  %
\IEEEoverridecommandlockouts      
\usepackage{cite}
\usepackage{amsmath,amssymb,amsfonts}
\usepackage{algorithmic}
\usepackage{graphicx}
\usepackage{algorithm,algorithmic}
\usepackage{soul} 
\usepackage{xcolor}
\usepackage{tikz} 
\usepackage{color}
\usepackage{graphicx}
\graphicspath{{figures/}{./}}
\usepackage{textcomp}
\usepackage{trimclip}
\usepackage[mathscr]{euscript}
\usepackage{array}
\usepackage{eqparbox}
\usepackage{url}
\usepackage{relsize}
\usepackage{dsfont}
\usepackage{matlab-prettifier}

\newtheorem{theorem}{Theorem}

\newtheorem{proposition}[theorem]{Proposition}

\newtheorem{definition}{Definition}
\newtheorem{corollary}[theorem]{Corollary}

\newcommand{\vsp}{\vspace{0.05in}}

\newcommand{\calG}{\mathcal G}
\newcommand{\calV}{\mathcal V}
\newcommand{\calE}{\mathcal E}

\def\BibTeX{{\rm B\kern-.05em{\sc i\kern-.025em b}\kern-.08em
    T\kern-.1667em\lower.7ex\hbox{E}\kern-.125emX}}
\begin{document}

\bstctlcite{}
% \title{On Synchronization of High-dimensional Higher-order Networks through Quantized Tensor Trains}
\title{Controllability and Observability of Temporal Hypergraphs}

\author{Anqi Dong, Xin Mao, and Can Chen
\thanks{Anqi Dong is with the Department of Mechanical and Aerospace Engineering, University of California, Irvine, Irvine, CA 92617, USA (e-mail: anqid2@uci.edu).}
\thanks{Xin Mao is with the School of Data Science and Society, University of North Carolina at Chapel Hill, Chapel Hill, NC 27599, USA (email: XXX@unc.edu)}
\thanks{Can Chen is with the School of Data Science and Society and the Department of Mathematics, University of North Carolina at Chapel Hill, Chapel Hill, NC 27599, USA (e-mail: canc@unc.edu).}%
\thanks{Anqi Dong and Xin Mao contributed equally to this work.}
}

\maketitle
%%%%%%%%%%%%%%%%%%%%%%%%%%%%%%%%%%%%%%%%%%%%%%%%%%%%%%%%%%%%%%%%%%%%%%%%%%%%%%%%
\begin{abstract}
Numerous complex systems, such as those arisen in ecological networks, genomic contact networks, and social networks, exhibit higher-order and time-varying characteristics, which can be effectively modeled using temporal hypergraphs.  However, analyzing and controlling temporal hypergraphs poses significant challenges due to their inherent time-varying and nonlinear nature, while most existing methods predominantly target static hypergraphs. In this article, we generalize the notions of controllability and observability to temporal hypergraphs by leveraging tensor and nonlinear systems theory. Specifically, we establish tensor-based rank conditions to determine the weak controllability and observability of temporal hypergraphs. The proposed framework is further demonstrated with synthetic and real-world examples.
%Finally, we demonstrate our framework with synthetic and real-world examples.

\end{abstract}


\section{Introduction}\label{sec:intro}

Hypergraphs generalize graphs by allowing hyperedges to connect arbitrary subsets of nodes, capturing higher-order relationships unambiguously \cite{berge1984hypergraphs,wolf2016advantages,gao2020hypergraph}. Numerous real-world complex systems can be naturally represented using hypergraphs, including ecological networks, genomic contact networks, chemical reaction networks, co-authorship networks, and film actor/actress networks \cite{chen2023survey}. For instance, in ecological networks, species interactions often occur in higher-order combinations, where the relationship between two species can be influenced by one or more additional species \cite{bairey2016high}. Increasing evidence has revealed that higher-order interactions play a significant role in the dynamical processes of ecological networks \cite{bairey2016high,gibbs2022coexistence}. Therefore, understanding the system-theoretic properties of hypergraphs such as controllability and observability becomes imperative for effectively managing and predicting the dynamics of complex systems. 

A variety of results have been developed concerning the dynamics of static hypergraphs. For instance, Chen et al. \cite{chen2021controllability} pioneered the development of a generalized Kalman's rank condition to determine the controllability of hypergraphs by leveraging homogeneous polynomial systems theory and tensor algebra, extending previous findings on graph controllability \cite{liu2011controllability,yuan2013exact}. Notably, the authors applied the rank condition to compute the minimum number of driver nodes of real-world hypergraphs. Subsequently, Pickard et al. \cite{pickard2023observability} employed a similar approach to investigate the weak observability of hypergraphs and the associated minimum number of sensor nodes. Recently, Zhang et al. \cite{zhang2024global} extended Pickard et al.'s work in a broader scope for hypergraph observability.

\begin{figure}[t]
    \centering
    \includegraphics[width=\columnwidth,trim={0cm 7cm 0cm 5cm},clip]{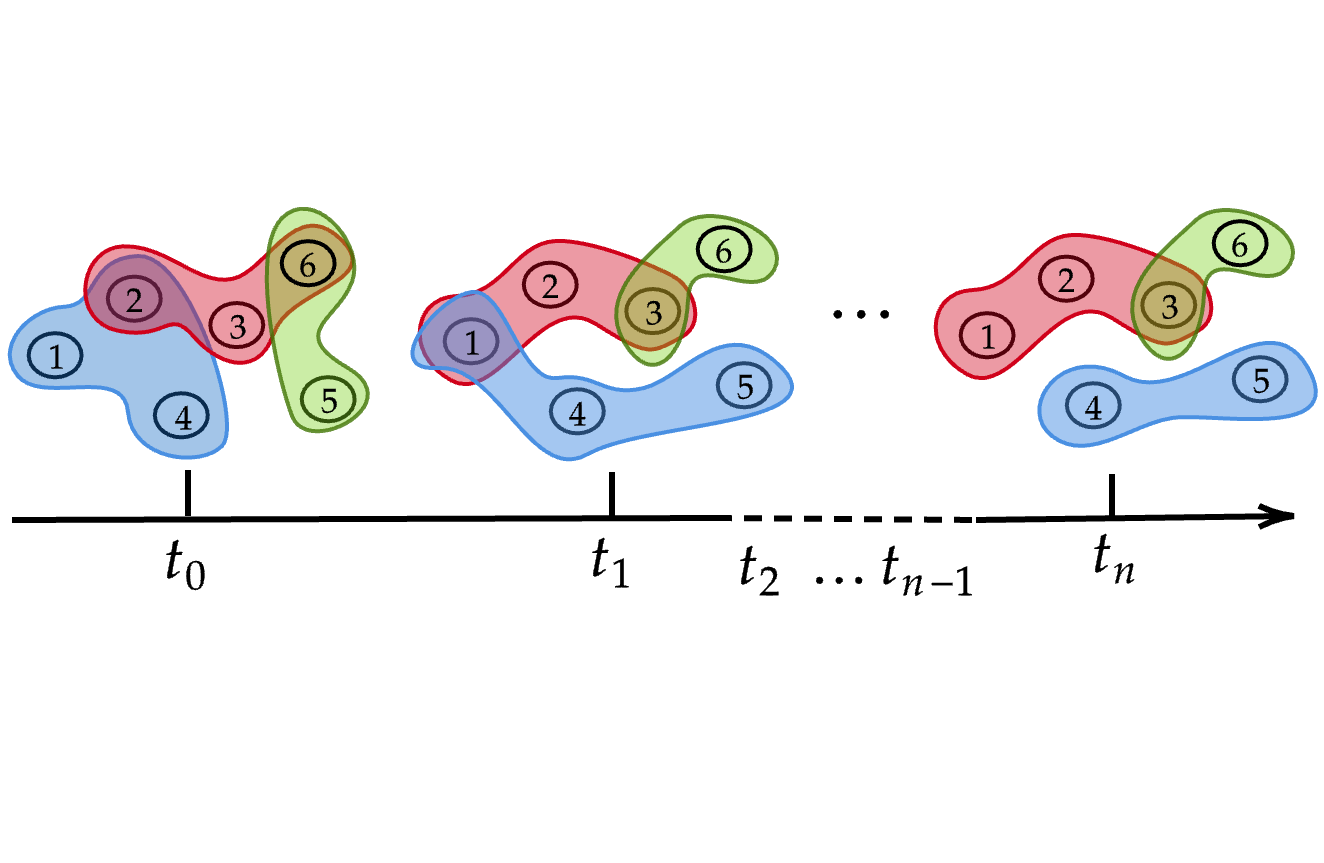}
    \caption{Temporal hypergraph whose structure changes over time. The connectivity is shown by the hyperedges represented by areas with distinct colors, which changes over time ($t_0,t_1,\dots, t_n$).}
    \label{fig:temp_graph}
    \vspace{-10pt}
\end{figure}

However, many real-world networks are time-varying and can be more accurately modeled using temporal hypergraphs, whose structure changes over time \cite{neuhauser2021consensus,myers2023topological} (see Fig. \ref{fig:temp_graph}). As an illustrative example, in food webs, species interactions and predator-prey relationships exhibit seasonal fluctuations, such as changes in bird migration patterns affecting prey populations \cite{schoenly1991temporal,closs1994spatial}. Moreover, these relationships can vary in response to environmental factors like temperature fluctuations, alterations in water flow patterns, and shifts in nutrient availability (e.g., warmer temperatures might favor certain fish species over others) \cite{schoenly1991temporal, closs1994spatial}. Consequently, understanding these temporal variations is essential for studying system properties and designing optimal control strategies for food webs, which has implications for species coexistence, biodiversity, and community persistence. Yet, most existing methods primarily focus on static hypergraphs, highlighting the need for the development of computational tools to analyze and control temporal hypergraphs.

In this article, we aim to extend previous efforts on network controllability and observability to temporal hypergraphs by exploiting tensor theory and nonlinear systems theory. The key contributions are listed as follows:
\begin{itemize}
    \item Building on \cite{chen2021controllability}, we employ a tensor-based time-varying polynomial system model to capture the dynamics of temporal hypergraphs.
    \item We establish tensor-based rank conditions to determine the weak controllability and observability of time-varying polynomial systems and temporal hypergraphs.
    \item We demonstrate our framework with synthetic temporal hypergraphs and real-world ecological networks. 
\end{itemize}

The article is organized into six sections. In Section \ref{sec:prelim}, we briefly review the notions of hypergraphs, tensor-based polynomial systems, and nonlinear controllability and observability. We derive tensor-based rank criteria to assess the weak controllability and observability of temporal hypergraphs in Sections \ref{sec:contr} and \ref{sec:obser}, respectively. Two numerical examples are provided in Section \ref{sec:example},  and we conclude with future directions in Section \ref{sec:conc}. Throughout this article, we denote scalars by standard lowercase characters, vectors by bold lowercase characters, matrices by bold uppercase characters, and tensors by script uppercase characters.

\section{Preliminaries}\label{sec:prelim}

\subsection{Hypergraphs}
An unweighted hypergraph $\calG = \{\calV, \calE \}$ can be characterized by a doublet, where $\calV=\{v_1,v_2,\dots,v_n\}$ is the node set and $\calE =\{e_1,e_2,\dots,e_m\}$ is the hyperedge set such that $e_p\subseteq \calV$ for $p=1,2,\dots,m$ \cite{berge1984hypergraphs}. Two nodes are called adjacent if they belong to the same hyperedge. The degree of a node is equal to the number of hyperedges that contain that node. If all hyperedges contain exactly $k$ nodes, $\calG$ is called a $k$-uniform hypergraph (note that traditional graphs are therefore 2-uniform hypergraphs). Significantly, every $k$-uniform hypergraph with $n$ nodes can be represented as a $k$th-order, $n$-dimensional symmetric tensor (i.e., invariant under any permutation of its indices) \cite{chen2021controllability}.

\vsp
\begin{definition}[Adjacency tensors]
    Suppose that $\calG$ is a $k$-uniform hypergraph with $n$ nodes. The adjacency tensor $\mathscr{A}\in\mathbb{R}^{n\times n\times \stackrel{k}{\cdots}\times n}$ of $\calG$ is defined as
    \begin{equation}
        \mathscr{A}_{i_1i_2\cdots i_k}=
        \begin{cases}
            1/(k-1)! \quad \text{if } (i_1,i_2,\dots,i_k)\in \calE\\
            0 \quad \text{otherwise}
        \end{cases},
    \end{equation}
where $(k-1)!$ denotes the factorial of $(k-1)$, such that the degree of node $i$ is computed as 
\begin{equation*}
    d_i=\sum_{i_2=1}^n\sum_{i_3=1}^n\cdots\sum_{i_k=1}^n \mathscr{A}_{i_1i_2\cdots i_k}.
\end{equation*}
\end{definition}
\vsp

For non-uniform hypergraphs, we can represent them by combining adjacency tensors of different orders. A temporal hypergraph is a generalization of a (static) hypergraph that incorporates a time dimension \cite{neuhauser2021consensus,myers2023topological}, making the adjacency tensors time-dependent, i.e., $\mathscr{A}_j(t)$ for $j=2,3,\dots,k,$ where $k$ the maximum cardinality of the hyperedges.

\subsection{Tensor-based Polynomial Systems}
The dynamics of hypergraphs can be modeled using tensor-based polynomial systems \cite{chen2021controllability, zhang2024global}. First, we introduce the operation of tensor matrix multiplications. Given a tensor $\mathscr{A}\in\mathbb{R}^{n\times n\times \stackrel{k}{\cdots}\times n}$, the tensor matrix multiplication $\mathscr{A}\times_p \textbf{M}$ along mode $p$ for a matrix $\textbf{M}\in\mathbb{R}^{n\times m}$ is defined as
\begin{equation*}
    (\mathscr{A}\times_p\textbf{M})_{i_1i_2\cdots i_{p-1}ji_{p+1}\cdots i_k}  =\sum_{i_p=1}^n \mathscr{A}_{i_1i_2\cdots i_k}\textbf{M}_{i_pj}.
\end{equation*}
When $m=1$, the definition reduces to the tensor vector multiplication. Analogous to graph dynamics with linear systems \cite{liu2011controllability}, the dynamics of a hypergraph can be naturally represented using tensors.

\vsp
\begin{definition}[Hypergraph dynamics]
Suppose that $\calG$ is a hypergraph with $n$ nodes and maximum hyperedge cardinality $k$. The dynamics of $\calG$ can be represented as
\begin{equation}\label{eq:hdyn}
    \dot{\textbf{x}}(t) = \sum_{j=2}^k\mathscr{A}_j\times_1 \textbf{x}(t)\times_2\textbf{x}(t)\times_3\cdots\times_{j-1}\textbf{x}(t),
\end{equation}
where $\mathscr{A}_j\in\mathbb{R}^{n\times n\times \stackrel{j}{\cdots}\times n}$ are the $j$th-order adjacency tensors of $\calG$ and $\textbf{x}(t)\in\mathbb{R}^n$ represents the state of each node. For simplicity, we can rewrite \eqref{eq:hdyn} as $\dot{\textbf{x}}(t) =\sum_{j=2}^k \mathscr{A}_j\textbf{x}(t)^{j-1}$.
\end{definition}
\vsp

The tensor-based dynamical system \eqref{eq:hdyn} in fact belongs to the family of polynomial systems of degree $k-1$, which have been used to capture the dynamics of higher-order interactions in various fields \cite{bairey2016high,fujarewicz2005fitting,tang2022optimizing}.  Similarly, the dynamics of a temporal hypergraph can be thus represented using a tensor-based time-varying polynomial system, i.e., 
\begin{equation}\label{eq:temp_system}
    \dot{\textbf{x}}(t) = \sum_{j=2}^k\mathscr{A}_j(t)\textbf{x}(t)^{j-1},
\end{equation}
where $\mathscr{A}_j(t)$ are the time-dependent $j$th-order adjacency tensors of the hypergraph. Our goal in this article is to analyze the controllability and observability of the tensor-based time-varying polynomial system \eqref{eq:temp_system}. 

\subsection{Nonlinear Controllability \& Observability}
Controllability and observability are two of the most fundamental system-theoretic properties of a dynamical system. Initially introduced for linear systems, both concepts can be verified using the classical Kalman's rank condition \cite{kalman1963mathematical}. However, defining and verifying the two properties becomes significantly more complex for nonlinear systems \cite{hermann1977nonlinear}.

\vsp
\begin{definition}[Local weak  controllability]
    A nonlinear control system is called locally weakly controllable at  $\textbf{x}_0$ if, for any state $\textbf{x}_1$ within a small neighborhood of $\textbf{x}_0$, there exists a piecewise continuous control function that drives the system from $\textbf{x}_0$ to $\textbf{x}_1$ within a finite time interval. 
\end{definition}
\vsp
\begin{definition}[Local weak observability]
    A nonlinear output system is called locally weakly observable at $\textbf{x}_0$ if, there exists a neighborhood of $\textbf{x}_0$ such that if two initial states  $\textbf{x}_0$ and $\textbf{x}_1$ (within the neighborhood) produce indistinguishable outputs over a finite time interval, then $\textbf{x}_0=\textbf{x}_1$.
\end{definition}
\vsp

Weak controllability and observability refer to the cases when both definitions hold for any $\textbf{x}_0\in\mathbb{R}^n$. For nonlinear systems including homogeneous systems, many conditions have been established to determine local weak controllability and observability \cite{hermann1977nonlinear,baillieul1981controllability,carravetta2019test,martinelli2020rank,martinelli2022extension}. 

\section{Controllability}\label{sec:contr}
In this section, we propose a tensor-based rank condition for assessing the weak controllability of tensor-based polynomial time-varying  systems of degree $k-1$ with linear inputs, i.e., 
\begin{equation}\label{eq:hgdl}
    \dot{\textbf{x}}(t)=\sum_{j=2}^k\mathscr{A}_j(t)\textbf{x}(t)^{j-1} + \textbf{B}(t)\textbf{u}(t),
\end{equation}
where $\mathscr{A}_j(t)\in\mathbb{R}^{n\times n\times \stackrel{j}{\cdots}\times n}$ are the $j$th-order time-dependent dynamic tensors, $\textbf{B}(t)\in\mathbb{R}^{n\times m}$ is the time-dependent control matrix, and $\textbf{u}(t)\in\mathbb{R}^{m}$ is the control input. Additionally, we introduce the operator of Lie brackets from differential geometry, which plays a crucial role in proving controllability.
For more details of Lie brackets, we refer to \cite{bloch2015nonholonomic}.

\begin{definition}[Lie brackets]
Given two vector fields \textbf{f} and \textbf{g}, the Lie bracket of \textbf{f} and \textbf{g} at a point \textbf{x} is defined as
\begin{equation}\label{eq:lieb}
    [\textbf{f}, \textbf{g}]_{\textbf{x}} =\nabla \textbf{g}(\textbf{x})\textbf{f}(\textbf{x})-\nabla \textbf{f}(\textbf{x})\textbf{g}(\textbf{x}),
\end{equation}
where $\nabla$ denotes the gradient operation. 
\end{definition}

\subsection{Tensor-based Rank Condition}
We assume $\mathscr{A}_j(t)$ to be symmetric as they represent the adjacency tensors of temporal hypergraphs of different orders. The results for non-symmetric dynamic tensors can be extended in a similar manner. We formulate the tensor-based controllability rank condition as follows.

% and define the operation $\mathscr{A}(t)\star\textbf{M}(t)$ as the matrix formed from
% \begin{equation*}
%   \mathscr{A}(t)\times_1 \textbf{m}_1(t)\times_2\textbf{m}_2(t)\times_3\dots\times_{k-1}\textbf{m}_{k-1}(t)
% \end{equation*}
% where $\textbf{m}_p(t)$ are drawn from the columns of $\textbf{M}(t)$  with repetition allowed.
\vsp
\begin{proposition}[Controllability]\label{prop:ctrb}
The tensor-based polynomial time-varying  control system \eqref{eq:hgdl} is locally weakly controllable if and only if the controllability matrix is defined as
    \begin{equation}\label{eq:ctrb}
        \textbf{C}(\textbf{x},t)=\begin{bmatrix}
            \textbf{M}_0(\textbf{x},t) & \textbf{M}_1(\textbf{x},t) & \cdots & \textbf{M}_{n-1}(\textbf{x},t)
        \end{bmatrix},
    \end{equation}
where $\textbf{M}_0(\textbf{x},t) = \textbf{B}(t)$, and 
    \begin{align*}
        \textbf{M}_i(\textbf{x},t) &= \Bigg[\sum_{j=2}^k(j-1)
            \mathscr{A}_j(t)\textbf{x}^{j-2}\textbf{M}_{i-1}(\textbf{x},t)-\frac{\partial \textbf{M}_{i-1}(\textbf{x},t)}{\partial t} \\ & \quad \quad-\frac{\partial \textbf{m}^{(1)}_{i-1}(\textbf{x},t)}{\partial \textbf{x}}\textbf{B}(t)\text{ }\cdots \text{ } -\frac{\partial \textbf{m}^{(l)}_{i-1}(\textbf{x},t)}{\partial \textbf{x}}\textbf{B}(t)\Bigg],
    \end{align*}
 for $i=1,2,\dots,n-1$, has full rank. Here, $\textbf{m}_{i-1}^{(p)}(\textbf{x},t)$ denotes the $p$th column of $\textbf{M}_{i-1}(\textbf{x},t)$ with total {$l$ columns.}
\end{proposition}
\vsp
\begin{proof}
Based on nonlinear systems theory \cite{bloch2015nonholonomic,martinelli2020rank}, the controllability matrix (distribution) can be computed by recursively evaluating the Lie brackets of $\{\textbf{B}(t),\sum_{j=2}^k\mathscr{A}_j(t)\textbf{x}^{k-1}\}$, treating $t$ as another state. Without loss of generality, assume that $m=1$, i.e., $\textbf{B}(t)=\textbf{b}(t)\in\mathbb{R}^n$. Since $\mathscr{A}(t)$ is symmetric, for each iteration, it follows that
\begin{align*}
    \textbf{M}_i &=
        \Bigg[\langle \textbf{m}_{i-1}^{(1)}, \sum_{j=2}^k\mathscr{A}_j(t)\textbf{x}^{j-1}\rangle_{\textbf{x}}  \text{ }\cdots \text{ }  \langle\textbf{m}_{i-1}^{(l)}, \sum_{j=2}^k\mathscr{A}_j(t)\textbf{x}^{j-1}\rangle_{\textbf{x}}\\
        & \quad \quad [\textbf{m}_{i-1}^{(1)}, \textbf{b}(t)]_{\textbf{x}} \text{ }\cdots \text{ } [\textbf{m}_{i-1}^{(l)}, \textbf{b}(t)]_{\textbf{x}} \Bigg],
\end{align*}
for some $l$, where
\begin{equation*}
    \langle \textbf{f},\textbf{g}\rangle_\textbf{x}=[\textbf{f},\textbf{g}]_\textbf{x}-\frac{\partial \textbf{f}}{\partial t}.
\end{equation*}
Using the properties of tensor vector multiplications and Lie brackets, these brackets can be computed as 
{\small
\begin{align*}
    \langle \textbf{m}_{i-1}^{(p)}, \sum_{j=2}^k\mathscr{A}_j(t)\textbf{x}^{j-1}\rangle_{\textbf{x}}
    & = \sum_{j=2}^k \langle \textbf{m}_{i-1}^{(p)},\mathscr{A}_j(t)\textbf{x}^{j-1}\rangle_{\textbf{x}}\\
    & = \sum_{j=2}^k(j-1)\mathscr{A}(t)\textbf{x}^{j-2}\textbf{m}_{i-1}^{(p)} - \frac{\partial \textbf{m}_{i-1}^{(p)}}{\partial t},\\
    [\textbf{m}_{i-1}^{(p)}, \textbf{b}(t)]_{\textbf{x}} &= -\frac{\partial \textbf{m}^{(p)}_{i-1}}{\partial \textbf{x}}\textbf{b}(t).
\end{align*}}%
The recursive process will converge at most $n-1$ steps, and the result follows immediately. 
% Based on the findings in \cite{martinelli2020rank}, we can compute the controllability matrix by recursively evaluating the Lie brackets of $\{\textbf{B}(t),\mathscr{A}(t)\textbf{x}^{k-1}\}$. Without loss of generality, assume that $m=1$, i.e., $\textbf{B}(t)=\textbf{b}(t)\in\mathbb{R}^n$. Since $\mathscr{A}(t)$ is symmetric, for each iteration, we have
% \begin{align*}
%     \textbf{M}_i &=
%         \Big[\langle \textbf{m}_{i-1}^{(1)}, \mathscr{A}(t)\textbf{x}^{k-1}\rangle  \text{ }\cdots \text{ }  \langle\textbf{m}_{i-1}^{(l)}, \mathscr{A}(t)\textbf{x}^{k-1}\rangle\\
%         & \quad \quad [\textbf{m}_{i-1}^{(1)}, \textbf{b}(t)] \text{ }\cdots \text{ } [\textbf{m}_{i-1}^{(l)}, \textbf{b}(t)] \Big],
% \end{align*}
% for some $l$, where $[\textbf{f},\textbf{g}]$ denotes the Lie bracket, and $$
% \langle \textbf{f},\textbf{g}\rangle=[\textbf{f},\textbf{g}]-\frac{\partial \textbf{f}}{\partial t}.
% $$ 
% According to the properties of tensor vector multiplications, these brackets can be computed as 
% \begin{align*}
%     \langle \textbf{m}_{i-1}^{(p)}, \mathscr{A}(t)\textbf{x}^{k-1}\rangle &= (k-1)\mathscr{A}(t)\textbf{x}^{k-2}\textbf{m}_{i-1}^{(p)} - \frac{\partial \textbf{m}_{i-1}^{(p)}}{\partial t},\\
%     [\textbf{m}_{i-1}^{(p)}, \textbf{b}(t)] &= -\frac{\partial \textbf{m}^{(p)}_{i-1}}{\partial \textbf{x}}\textbf{b}(t).
% \end{align*}
% Since we are dealing with a finite vector space, the recursive process will converge at most $n-1$ steps, and the result follows immediately.
\end{proof}
\vsp

The controllability matrix $\textbf{C}(\textbf{x},t)$ in \eqref{eq:ctrb} is in general state- and time-dependent. If the rank condition holds for all $\textbf{x}=\textbf{x}_0\in\mathbb{R}^n$ and $t=t_0\in\mathbb{R}^{+}$, the tensor-based time-varying polynomial system with linear inputs \eqref{eq:hgdl} is weakly controllable. To compute the rank of the controllability matrix, we can utilize either numerical or symbolic computations. Numerical computations involve evaluating the controllability matrix at a finite number of sample points $(\textbf{x},t)$ and computing the rank using numerical rank operations. Symbolic computations entail expressing the entries of the controllability matrix as symbolic functions of \textbf{x} and $t$. By leveraging symbolic algebra software (e.g., MATLAB Symbolic Toolbox), we can perform exact calculations to determine the rank of the controllability matrix. However, symbolic computations can be computationally intensive and may struggle with the complexity for high-dimensional systems, potentially leading to errors.

For $k$-uniform temporal hypergraphs where the underlying dynamics is homogeneous of degree $k-1$, i.e., $j=k$ only in \eqref{eq:hgdl}, we can obtain a simpler rank condition. In particular, when $k=2$, it reduces to traditional temporal graphs with linear time-varying dynamics.  

\vsp
\begin{corollary}[Homogeneous case]\label{coro:homo}
    The tensor-based homogeneous time-varying  control system \eqref{eq:hgdl} with $j=k$ only is locally weakly controllable if and only if the controllability matrix \eqref{eq:ctrb} with $\textbf{M}_0(\textbf{x},t) = \textbf{B}(t)$ and 
        \begin{align*}
        \textbf{M}_i(\textbf{x},t) &= \Bigg[
            (k-1)\mathscr{A}_k(t)\textbf{x}^{k-2}\textbf{M}_{i-1}(\textbf{x},t)-\frac{\partial \textbf{M}_{i-1}(\textbf{x},t)}{\partial t} \\ & \quad \quad-\frac{\partial \textbf{m}^{(1)}_{i-1}(\textbf{x},t)}{\partial \textbf{x}}\textbf{B}(t)\text{ }\cdots \text{ } -\frac{\partial \textbf{m}^{(l)}_{i-1}(\textbf{x},t)}{\partial \textbf{x}}\textbf{B}(t)\Bigg],
    \end{align*}
for $i=1,2,\dots,n-1$, has full rank.
\end{corollary}
\vsp
\begin{proof}
    Due to the bilinearity of Lie brackets \eqref{eq:lieb}, the result follows directly by eliminating the summation.
\end{proof}

\vsp
\begin{corollary}[Linear case]
    The linear time-varying  control system \eqref{eq:hgdl} with $k=2$ is  weakly controllable if and only if the controllability matrix \eqref{eq:ctrb} with $\textbf{M}_0(\textbf{x},t) = \textbf{B}(t)$ and 
        \begin{align*}
        \textbf{M}_i(t) = 
            \mathscr{A}_2(t)\textbf{M}_{i-1}(t)-\frac{\partial \textbf{M}_{i-1}(\textbf{x},t)}{\partial t}
    \end{align*}
 for $i=1,2,\dots,n-1$, has full rank.
\end{corollary}
\vsp
\begin{proof}
    The result follows by setting $k=2$ in $\textbf{M}_i(\textbf{x},t)$ from Proposition \ref{prop:ctrb}, where it becomes state-independent.
\end{proof}
\vsp
Significantly, the condition above aligns with the controllability rank condition proposed in \cite{silverman1967controllability} for linear time-varying control systems. In fact, we can even drop the term ``weakly'' as it is proven for full controllability.

% Significantly, when $k=2$, the result reduces to the controllability rank condition for linear time-varying systems with linear inputs and holds globally. 

% \vsp
% \begin{corollary}[Linear case]
%     The control system \eqref{eq:hgdl} with $k=2$ (i.e., linear time-varying systems) is weakly controllable if and only if the controllability matrix defined as
%     \begin{equation*}
%         \textbf{C}(t)=\begin{bmatrix}
%             \textbf{M}_0(t) & \textbf{M}_1(t) & \cdots & \textbf{M}_{n-1}(t)
%         \end{bmatrix},
%     \end{equation*}
%     where $\textbf{M}_0(t) = \textbf{B}(t)$, and 
%     \begin{align*}
%         \textbf{M}_i(t) = 
%             \mathscr{A}(t)\textbf{M}_{i-1}(t)-\frac{\partial \textbf{M}_{i-1}(\textbf{x},t)}{\partial t}
%     \end{align*}
%  for $i=1,2,\dots,n-1$, has full rank.
% \end{corollary}
% \vsp
% \begin{proof}
%     The result follows by setting $k=2$ in $\textbf{M}_i(\textbf{x},t)$ from Proposition \ref{prop:ctrb}, where it becomes state-independent.
% \end{proof}

% The condition aligns with the one proposed in \cite{silverman1967controllability} for linear time-varying control systems. In fact, we can even drop the term ``weakly'' as it is proven for full controllability \cite{silverman1967controllability}.

\subsection{Controllability of Temporal Hypergraphs}

In network science, the notion of the minimum number of driver nodes (MNDN), introduced by Liu et al. \cite{liu2011controllability}, represents the smallest set of nodes necessary to fully control an entire network. This idea has broad applications, from understanding the behavior of biological systems \cite{angulo2019theoretical} to optimizing control strategies in social networks for disease prevention \cite{danon2011networks}. By leveraging Proposition \ref{prop:ctrb} and Corollary \ref{coro:homo}, we can discuss the controllability of temporal hypergraphs.  Similar to the approach used in \cite{chen2021controllability}, we aim to identify the MNDN of a temporal hypergraph such that the corresponding controllability matrix of the underlying dynamics has full rank. The MNDN is a powerful concept that can be used to steer the dynamics of temporal hypergraphs with minimal effort \cite{liu2011controllability}. It can also serve as an indication of the robustness of temporal hypergraphs \cite{chen2021controllability}. Intuitively, if the MNDN of a temporal hypergraph is high, it will require more effort or energy to control the hypergraph or steer the underlying system. Furthermore, the MNDN provides insights into the underlying topology of the temporal hypergraph, which helps identify key nodes that play a crucial role in the system's dynamics. For simplicity, we assume that the control matrix is time-independent (i.e., \textbf{B}) and each input can only be imposed on one node (i.e., the columns of \textbf{B} are the scaled standard basis vectors).

Identifying the MNDN of a temporal hypergraph through a brute-force search is NP-hard and time-consuming  \cite{chen2021controllability}. We offer a simple heuristic method for approximating the minimum subset of driver nodes in a temporal hypergraph, where nodes are selected based on the maximum change in the rank of the controllability matrix (Algorithm \ref{alg:1}). The rank computations in Step 6 can be performed either numerically or symbolically. In Step 7, if multiple $v^*$ are obtained, we can either pick one randomly or break the tie based on their degrees. Note that the final subset of driver nodes $\mathcal{D}$ is highly likely to be minimal (i.e., MNDN), but it is not guaranteed.

\begin{algorithm}[t] 
\caption{Greedy driver nodes selection}
\label{alg:1}
\begin{algorithmic}[1]
\STATE{Given symmetric time-varying adjacency tensors $\mathscr{A}_j(t)\in\mathbb{R}^{n\times n\times \stackrel{j}{\cdots} \times n}$ of a temporal hypergraph for $j=2,\dots,k$}\\
\STATE{Let $\mathcal{V}=\{1,2,\dots,n\}$ and the index set $\mathcal{D}=\emptyset$}\\
\STATE{Let $\textbf{C}_\mathcal{D}(\textbf{x},t)$ be the controllability matrix with the control matrix formed from the index set $\mathcal{D}$}
\WHILE{$\text{rank}(\textbf{C}_\mathcal{D}(\textbf{x},t))<n$}
\FOR{$v\in \mathcal{V}\setminus \mathcal{D}$}
\STATE{Compute 
\begin{equation}\label{eq:deltarank}
    \Delta(v)=\text{rank}(\textbf{C}_{\mathcal{D}\cup \{v\}}(\textbf{x},t))-\text{rank}(\textbf{C}_\mathcal{D}(\textbf{x},t))
\end{equation}
}\\
\ENDFOR
\STATE{Set $v^{*} = \text{argmax}_{v\in \mathcal{V}\setminus \mathcal{D}}\Delta(v)$}\\
\STATE{Set $\mathcal{D}=\mathcal{D}\cup \{v^*\}$}
\ENDWHILE
\RETURN subset of driver nodes $\mathcal{D}$
\end{algorithmic}
\end{algorithm}

\section{Observability} \label{sec:obser}
In this section, we propose a tensor-based rank condition for assessing the weak observability of tensor-based polynomial time-varying  systems of degree $k-1$ with linear outputs, i.e., 
\begin{align}\label{eq:hgdo}
\begin{cases}
    \dot{\textbf{x}}(t)=\sum_{j=2}^k\mathscr{A}_j(t)\textbf{x}(t)^{j-1}  \\
	\textbf{y}(t)=\textbf{L}(t)\textbf x(t)
\end{cases},
\end{align}
where $\textbf{L}(t)\in\mathbb{R}^{q\times n}$ is the time-dependent output matrix, and $\textbf{y}\in\mathbb{R}^q$ is the output. Again, we assume that $\mathscr{A}_j(t)$ are symmetric. Additionally, we provide a brief review of the Lie derivative operator, which plays a crucial role in proving observability. 
\begin{definition}[Lie derivatives]
    Given a vector field \textbf{f} and a scalar field $h$, the Lie derivative of $h$ along \textbf{f} is defined as
    \begin{equation}
        \mathcal{L}_{\textbf{f}}h = (\partial h/\partial \textbf{x})\textbf{f}.
    \end{equation}
    Moreover, it satisfies
$
    \mathcal{L}_{\textbf{f}} Dh = D\mathcal{L}_{\textbf{f}}h
$
where $D$ denotes the differential operator. 
\end{definition}

Detailed definitions and properties of Lie derivatives can be found in \cite{bloch2015nonholonomic}. We formulate the tensor-based observability rank condition as follows.

\vsp
\begin{proposition}[Observability]\label{prop:2}
The tensor-based polynomial time-varying  output system \eqref{eq:hgdo} is locally weakly observable if and only if the observability matrix that defined as 
\begin{equation}\label{eq:otrb}
\textbf{O}(\textbf{x},t)=\left(\begin{matrix}\textbf N_0(\textbf{x},t)\\\textbf N_1(\textbf{x},t)\\\vdots\\\textbf N_{n-1}(\textbf{x},t)\end{matrix}\right),
\end{equation}
where $\textbf{N}_0\left(\textbf{x},t)=D(\textbf{L}(t)\textbf{x}\right)$, and 
\[
\mathbf{N}_{i}(\textbf{x},t)=\sum_{j=2}^k\frac{\partial\textbf{N}_{i-1}(\textbf{x},t)}{\partial \textbf{x}}\mathscr{A}_j(t)\textbf{x}^{j-1}+\frac{\partial\textbf{N}_{i-1}(\textbf{x},t)}{\partial t}
\] 
for $i=1,2,\dots,n-1$, has full rank.	
\end{proposition}
\vsp
\begin{proof}
Based on nonlinear systems theory \cite{bloch2015nonholonomic,martinelli2022extension}, the observability matrix (codistribution) can be computed by evaluating the Lie derivatives of the output along the system state, treating $t$ as another state. Since $\mathscr{A}_j(t)$ are symmetric, for each iteration, it follows that
\begin{align*}
    \textbf{N}_i = \Big[ \tilde{\mathcal{L}}_{\sum_{j=2}^k\mathscr{A}_j(t)\textbf{x}^{j-1}} \textbf{n}^{(1)}_{i-1} \text{ }\cdots\text{ } \tilde{\mathcal{L}}_{\sum_{j=2}^k\mathscr{A}_j(t)\textbf{x}^{j-1}} \textbf{n}^{(l)}_{i-1}\Big],
\end{align*}
where $\textbf{n}_{i-1}^{(p)}$ is the $p$th row of $\textbf{N}_{i-1}$ (with total $l$ rows), and $$\tilde{\mathcal{L}}_{\textbf{f}}h=\frac{\partial h}{\partial t} + \frac{\partial h}{\partial \textbf{x}}\textbf{f}.$$ Using the properties of tensor vector
multiplications and Lie derivatives,  each element can be computed as
\begin{equation*}
    \tilde{\mathcal{L}}_{\sum_{j=2}^k\mathscr{A}_j(t)\textbf{x}^{j-1}} \textbf{n}^{(p)}_{i-1} = \sum_{j=2}^k\frac{\partial \textbf{n}_{i-1}^{(p)}}{\partial \textbf{x}}\mathscr{A}_j(t)\textbf{x}^{j-1} + \frac{\partial \textbf{n}_{i-1}^{(p)}}{\partial t}.
\end{equation*}
For example, $\mathbf{N}_{1}(\textbf{x},t)$ can be computed as
\begin{align*}
    \mathbf{N}_{1}(\textbf{x},t)&=
    D\Bigg(\sum_{j=2}^k\textbf{L}(t)\mathscr{A}_j(t)\textbf{x}^{j-1}+\frac{\partial (\textbf{L}(t)\textbf{x}}{\partial t}\Bigg).
\end{align*}
The recursive process will converge at most $n-1$ steps, and the result follows immediately. 
% where 
% 	Denote the operator $\tilde L_fh=\frac{\partial h}{\partial t}+L_fh$, where $L_fh=\langle dh,f\rangle$ is the Lie derivatives of $h$ along $f$. Then it holds that $\tilde L_fdh=d\tilde L_fh.$ According to \cite{martinelli2020rank},
% We can compute the derivatives of output along system state as follows to determine the observability matrix:
%  \begin{align*}
% \tilde L_f\textbf{N}_{i-1}=\frac{\partial\textbf{N}_{i-1}}{\partial t}+\frac{\partial\textbf{N}_{i-1}}{\partial x}\mathscr{A}(t)\textbf{x}^{k-1}.
%  \end{align*}
%  The recursive process will converge in an open and dense set in at most $k-1$ steps. The proof is completed.
\end{proof}
\vsp

Similar to the controllability matrix, the observability matrix $\textbf{O}(\textbf{x},t)$ in \eqref{eq:otrb} is state- and time-dependent. If the rank condition holds for all $\textbf{x}=\textbf{x}_0\in\mathbb{R}^n$ and $t=t_0\in\mathbb{R}^{+}$, the tensor-based time-varying polynomial system with linear outputs \eqref{eq:hgdo} is weakly observable. For $k$-uniform temporal hypergraphs, we can drop the summation in $\textbf{N}_i(\textbf{x},t)$.
\vsp
\begin{corollary}[Homogeneous case]\label{coro:obshomo}
    The tensor-based homogeneous time-varying  output system \eqref{eq:hgdo} with $j=k$ only is locally weakly observable if and only if the observability matrix \eqref{eq:otrb}
    with $\textbf{N}_0(\textbf{x},t)=D(\textbf{L}(t)\textbf{x})$, and 
\[
\mathbf{N}_{i}(\textbf{x},t)=\frac{\partial\textbf{N}_{i-1}(\textbf{x},t)}{\partial \textbf{x}}\mathscr{A}_k(t)\textbf{x}^{k-1}+\frac{\partial\textbf{N}_{i-1}(\textbf{x},t)}{\partial t},
\] 
for $i=1,2,\dots,n-1$, has full rank.
\end{corollary}

% Additionally, when $k=2$, the result reduces to the observability rank condition for linear time-varying systems with linear outputs and holds globally.

\vsp
\begin{corollary}[Linear case]
The linear time-varying output system \eqref{eq:hgdo} with $k=2$  is weakly observable if and only if the observability matrix \eqref{eq:otrb} with $\textbf{N}_0(t)=\textbf{L}(t)$ and 
\begin{equation*}
    \textbf{N}_i(t)=\textbf{N}_{i-1}(t)\mathscr{A}_2(t) +\frac{d \textbf{N}_{i-1}(t)}{dt},
\end{equation*}
for $i=1,2,\dots,n-1$, has full rank.
\end{corollary}
\vsp

Notably, the criterion above coincides with the observability rank condition proposed in \cite{silverman1967controllability} for linear time-varying output systems, which is in fact for full observability. Based on Proposition \ref{prop:2} and Corollary \ref{coro:obshomo}, we can discuss the observability of temporal hypergraphs. Specifically, we are interested in identifying the minimum number of sensor nodes (MNSN) required for a temporal hypergraph to ensure that the associated observability matrix has full rank. The MNSN is a critical notion that enables the reconstruction of the full internal state of a temporal hypergraph \cite{liu2013observability}. Additionally, it is essential for designing feedback control, relying on estimations of the plant state based solely on the plant output or the measurements collected from its sensors \cite{pickard2023observability}. For simplicity, we assume that the output matrix is time-independent (i.e., \textbf{L}) and each output can only be imposed on one node. Therefore, we can utilize an approach similar to Algorithm \ref{alg:1} to find the ``minimum'' set of sensor nodes of a temporal hypergraph.

\section{Numerical Examples}\label{sec:example}
We illustrate our framework with two numerical examples, focusing on the controllability of temporal hypergraphs (the observability part can be implemented in a similar manner). All computations were performed using MATLAB R2022b with the Symbolic Toolbox. The associated code can be found at \url{https://github.com/dytroshut/temp.graph}.

% We start with synthetic examples of 3-uniform hypergraph with four nodes, and a real-world application in ecological dynamics with six species is considered. 

%The controllability matrix $\textbf{C}(\textbf{x},t)$ are derived for both cases with the minimum number of driver nodes (MNDN) identified.

\subsection{Uniform Temporal Hypergraphs}
\begin{figure}[t]
    \centering
    \includegraphics[width=0.5\columnwidth]{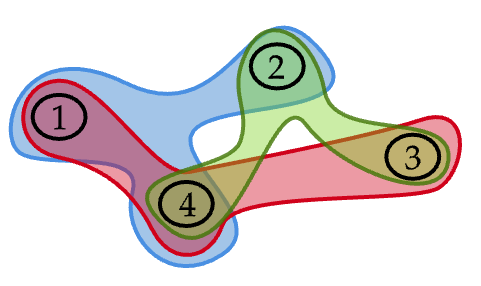}
    \caption{3-uniform hypergraph with time-varying weights on hyperedges.}
    \label{fig:example1}
\end{figure}

In this example, we constructed a synthetic  3-uniform temporal hypergraph with 4 nodes and three hyperedges $e_1 = \{v_1,v_2,v_4\}$, $e_2 = \{v_1,v_3,v_4\}$, and $e_3 = \{v_2,v_3,v_4\}$ (see Fig.~\ref{fig:example1}). All hyperedges have the same weight, which is set to $t$, i.e., $\mathscr{A}_{124}=\mathscr{A}_{134}=\mathscr{A}_{234}=t$ (and their corresponding permutations are also equal to $t$). We considered two different control matrices, which are defined as
\begin{align*}
    \textbf{B}_1(t) &= \begin{bmatrix}
        5t &2t &3t &t
    \end{bmatrix},\\
    \textbf{B}_2(t) &= \begin{bmatrix}
        \phantom{5}t &\phantom{2}t &\phantom{3}t &t
    \end{bmatrix}.
\end{align*}
Therefore, we computed the controllability matrices $\textbf{C}_{1}(\textbf{x},t)$ and $\textbf{C}_{2}(\textbf{x},t)$  based on Corollary~\ref{coro:homo} (due to the space limit, the two matrices are too large to display). Using the symbolic rank function, we determined that $\text{rank}(\textbf{C}_{1})=4$ and $\text{rank}(\textbf{C}_{2})=3$. Hence, the hypergraph dynamics with the control matrix $\textbf{B}_1(t)$ is weakly controllable. Additionally, while the symbolic rank of the controllability matrix $\textbf{C}_2(\textbf{x},t)$ is not full, the hypergraph dynamics with the control matrix $\textbf{B}_2(t)$ can achieve local weak controllability at certain $\textbf{x}\in\mathbb{R}^4$ within specific time intervals.

% The computation of $\textbf{C}_{1}(\textbf{x},t)$ and $\textbf{C}_{2}(\textbf{x},t)$ directly follows from Corollary~\ref{coro:homo}, with given corresponding control matrices $\textbf{B}_1$ and $\textbf{B}_2$ such that
% \begin{align*}
%     \textbf{B}_1(t) &= \begin{bmatrix}
%         5t &2t &3t &t
%     \end{bmatrix},\\
%     \textbf{B}_2(t) &= \begin{bmatrix}
%         \phantom{5}t &\phantom{2}t &\phantom{3}t &t
%     \end{bmatrix}.
% \end{align*}
% The rank of $\textbf{C}_{1}(\textbf{x},t)$ and $\textbf{C}_{2}(\textbf{x},t)$ can be thus obtained as $\text{rank}(\textbf{C}_{1})=4$ and $\text{rank}(\textbf{C}_{2})=3$, indicating the system with $\textbf{B}_1$ is fully controllable while not
% controllable with  $\textbf{B}_2$.

\begin{figure}[t]
    \centering
    \includegraphics[width=1\columnwidth]{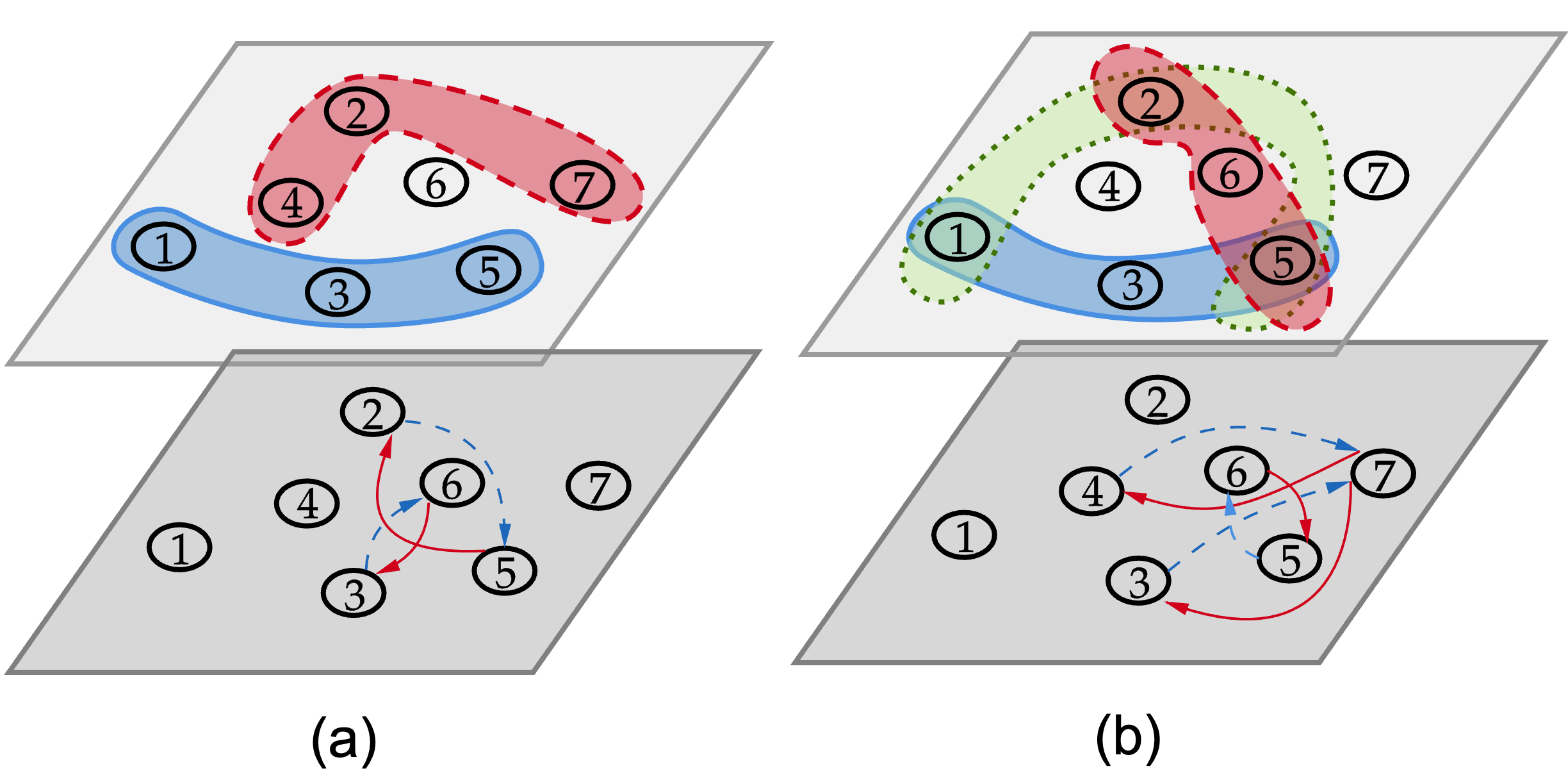}
    \caption{Two higher-order temporal ecological networks with pairwise and third-order interactions.}
    \label{fig:example2}
    \vspace{-10pt}
\end{figure}

\subsection{Higher-order Ecological Systems}
In ecological networks, species interactions are often time-varying and involve higher-order combinations. Understanding these interactions is crucial for accurately modeling ecosystem dynamics and predicting the impact of changes within the network. In our second example, we considered two temporal ecological networks with pairwise or third-order interactions among seven species $\mathcal{S} =\{s_1,s_2,\dots,s_7\}$ (see Fig. \ref{fig:example2}). The first network (a) involves pairwise interactions $\{s_2,s_5\}$ and $ \{s_3,s_6\}$ and third-order interactions $\{s_1,s_3,s_5\}$ and $\{s_2,s_4,s_7\}$.  The second network (b) consists of  pairwise interactions $\{s_3,s_7\}$, $\{s_4,s_7\}$, and $ \{s_5,s_6\}$ and third-order interactions  $\{s_1,s_3,s_5\}$, $\{s_2,s_5,s_6\}$, and $\{s_1,s_2,s_5\}$. We assumed that all interactions have the same weight $t$ in their adjacency matrices/tensors.

To determine the MNDN for both temporal hypergraphs, we followed the updating scheme outlined in Algorithm \ref{alg:1}, assuming that the control matrices are time-independent indicating the driver nodes. Therefore, the estimated MNDN of the two ecological networks can be computed as $\mathcal{D}_{1} =\{s_1,s_2,s_3\}$ and $ \mathcal{D}_{2} =\{s_3,s_7\}$, implying that the former ecological network structure is more robust to external inputs than the latter.  Note that the MNDN may not be unique. For instance, in the case of the second ecological network, alternative possible choices of MNDN include $\mathcal{D}_2=\{s_1,s_2\}$ and $\mathcal{D}_2=\{s_6,s_7\}$.

\section{Conclusion}\label{sec:conc}
In this article, we extended the framework of controllability and observability to temporal hypergraphs. By leveraging tensor and nonlinear systems theory, we developed novel tensor-based rank conditions to assess the weak controllability and observability of temporal hypergraphs. We further demonstrated the practical relevance and effectiveness of our approach using a synthetic example and real-world ecological networks. Incorporating temporal dynamics in complex systems provides new insights and computational tools for analyzing and controlling complex dynamics in time-varying higher-order networks. In the future, it would be valuable to explore stronger controllability and observability conditions for temporal hypergraphs and to apply this framework to large-scale, real-world hypergraph data (e.g., how to efficiently evaluate the symbolic rank for large-scale temporal hypergraphs?). Additionally, investigating optimal control design for temporal hypergraphs presents an important direction for further research. This could have significant implications for areas such as network stability, resource allocation, and dynamic behavior prediction in complex systems.

% \section*{References}
\bibliographystyle{IEEEtran}
\bibliography{references}

\end{document}